\documentclass[pra,twocolumn,showpacs,preprintnumbers,amssymb,tightenlines,epsfig,superscriptaddress]{revtex4}
\usepackage{graphicx}
\usepackage{color}
\usepackage[normalem]{ulem}

\usepackage{amsmath}
\usepackage[percent]{overpic}
\newcommand{\bra}[1]{\langle\,{#1}\, |}
\newcommand{\ket}[1]{|\,{#1}\,\rangle}

\newcommand{\rvec}[1]{{\mathbf{r}}}

\newcommand{\sub}[2]{{#1}_{\mbox{\!\! \scriptsize #2}}}

\def\beq{\begin{equation}}
\def\eeq{\end{equation}}

\def\CR{\nonumber\\[0.15cm]}
\newcommand{\id}{{1}}

\newcommand{\fref}[1]{Fig.~\ref{#1}}
\newcommand{\frefp}[2]{Fig.~\ref{#1}~(#2)}

\newcommand{\eref}[1]{Eq.~(\ref{#1})}
\newcommand{\esref}[2]{Eqs.~(\ref{#1}) and (\ref{#2})}

\newcommand{\sref}[1]{section~\ref{#1}}

\newcommand{\cref}[1]{chapter~\ref{#1}}

\newcommand{\Cref}[1]{Chapter~\ref{#1}}

\newcommand{\aref}[1]{appendix~\ref{#1}}
\newcommand{\bref}[1]{(\ref{#1})}

\usepackage{ulem}  
\definecolor{darkgreen}{rgb}{0,0.5,0}

\usepackage{bbold}
\usepackage{braket}
\usepackage{float}

\begin{document}

\title{Fano Resonances in Quantum Transport with Vibrations}
\author{Ajith Ramachandran}
\affiliation{Department of Physics, Indian Institute of Science Education 
and Research, Bhopal, Madhya Pradesh 462 066, India}
\author{Michael Genkin}
\affiliation{Max Planck Institute for the Physics of Complex Systems, N\"othnitzer Stra{\ss}e 38, D-01187 Dresden, Germany}
\author{Auditya Sharma}
\affiliation{Department of Physics, Indian Institute of Science Education 
and Research, Bhopal, Madhya Pradesh 462 066, India}
\author{Alexander Eisfeld}
\affiliation{Max Planck Institute for the Physics of Complex Systems, N\"othnitzer Stra{\ss}e 38, D-01187 Dresden, Germany}
\author{Sebastian W\"uster}
\email{sebastian@iiserb.ac.in}
\affiliation{Department of Physics, Indian Institute of Science Education 
and Research, Bhopal, Madhya Pradesh 462 066, India}
\author{Jan-Michael Rost}
\affiliation{Max Planck Institute for the Physics of Complex Systems, N\"othnitzer Stra{\ss}e 38, D-01187 Dresden, Germany}

\begin{abstract}
Quantum mechanical scattering involving continuum states coupled to a scatterer with a discrete spectrum gives rise to Fano resonances. Here we consider scatterers that possess internal vibrational degrees of freedom  in addition to discrete states.
Entanglement between the scattered excitation and vibrational modes complicates analytical and numerical calculations considerably.
For the example of one-dimensional scattering we develop a multichannel quantum scattering approach which can  determine  reflection and transmission probabilities in the presence of  vibrations.
Application to a linear chain coupled to a control unit containing vibrating sites shows that  vibrational degrees of freedom can have a profound effect on quantum transport. For suitable parameters, spectral regions which are opaque 
in the static case can be rendered transparent when vibrations are included. 
The formalism is general enough to be applicable to a variety of platforms for quantum transport including molecular aggregates, cold atom chains, quantum-dot  arrays and molecular wires based on conjugated polymers.
\end{abstract} 
\maketitle

\section{Introduction} 

Quantum transport of excitation, energy and entanglement are  fundamental features of a wide range of  systems ranging from quantum aggregates of organic molecules \cite{davydov1964theory,haken1972coupled,scherer1984exciton},  photosynthetic complexes \cite{van2000photosynthetic,kuhn1997pump,rebentrost2009environment} and cold atoms \cite{robicheaux2004simulation,ates2008motion,mulken2007survival,wuat+10,barredo:trimeragg} to quantum dot assemblies \cite{scholes2011excitons}. Such transport systems usually involve a regular part enabling
 wave-like transport  with continuous wave numbers and may contain a second part consisting of small subunits with a discrete spectrum which couple to the regular part and affect the transport. Generically, this scenario  gives rise to 
Fano resonances, ``bound states in the continuum'' originally described as ``strange discrete eigenvalues'' by von Neumann and Wigner in 1929 \cite{newi29a}.
They cause characteristic asymmetric features in the transport spectrum \cite{chakrabarti2007fano,deo1995asymmetric,miroshnichenko2005engineering}, also known as Fano profiles. The resonances even lead to complete reflection or complete transmission at certain resonant energies,  a useful resource for control and switching applications \cite{miroshnichenko2010fano,ismael2017connectivity}.  

Apart from Fano resonances, transport is often affected by internal vibrations or phonons, e.g., electron-phonon coupling in long conjugated molecules  \cite{heeger2001nobel,heeger1988solitons} or molecular aggregates \cite{saikin:excitonreview}. While mostly seen as impeding transport but inevitable, 
in Rydberg aggregates atomic motion can also induce quantum
transport \cite{wuat+10,wuat+11,mowu+11,moge+13a}. Here, we demonstrate how atomic motion
can serve to create a switch in transport systems. More specifically, we investigate
quantum transport along the regular part of the system beyond a localized ``obstacle'',
a subunit with one  dominant vibrational degree of freedom. We will see that the reflection and transmission profiles based on Fano resonances for a transport system with a {\em static} subunit are qualitatively altered if the latter can be  vibrationally excited. Thereby,  Fano resonances can be put to work for sensing and switching in transport transport systems even in the presence of directed or thermal motion.

\begin{figure}[htb]
\includegraphics[width=\columnwidth]{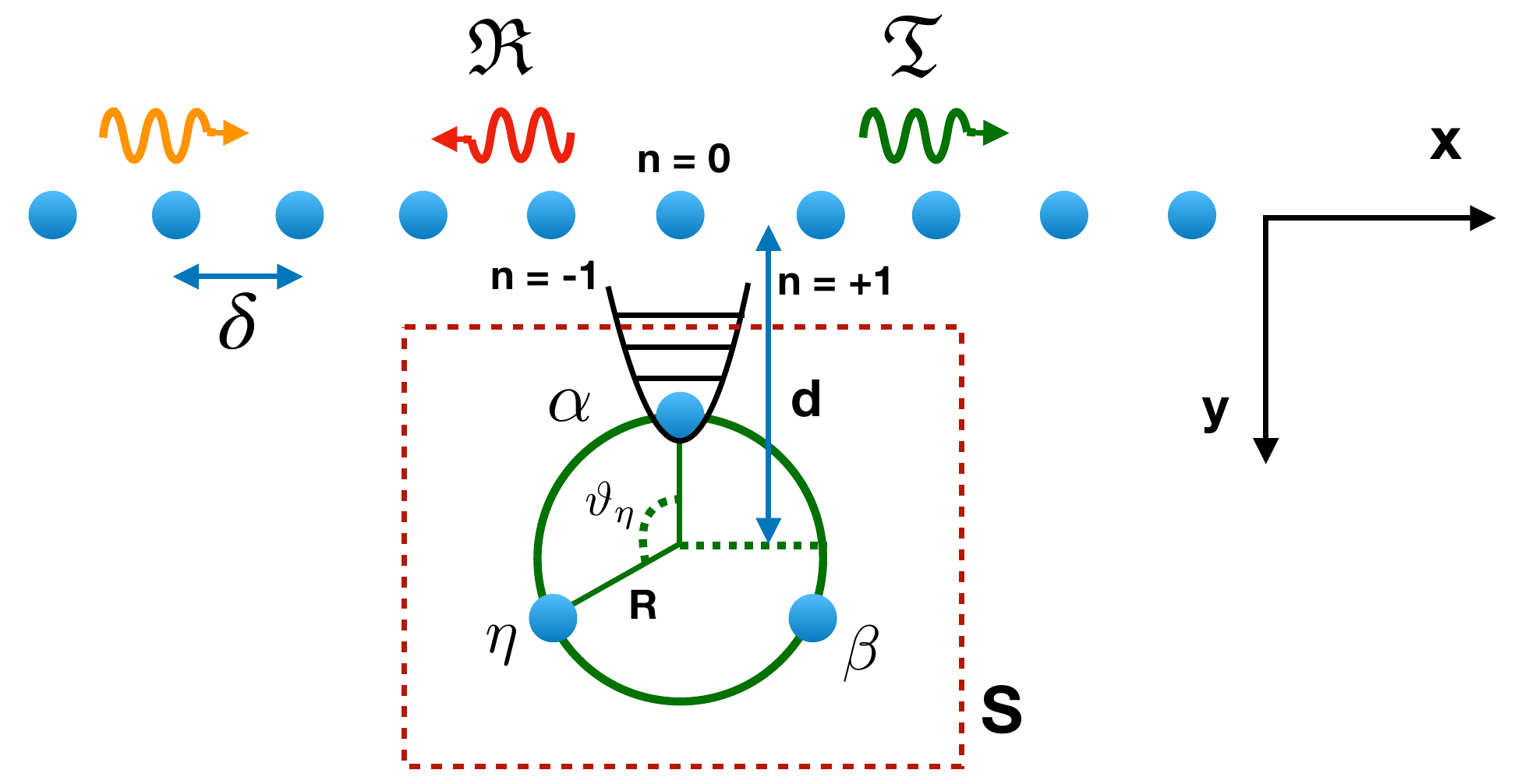}
\caption{\label{sketch} Sketch of the Fano-Anderson chain with vibrating elements. $N$  static monomers are arranged in an equidistant linear chain and three monomers $\alpha,~\beta,~\eta$ are arranged within a control unit $U$ confined to a ring of radius $R$. The center of the ring is at a distance $d$ from the chain.  The angular coordinates of the monomers on the ring are $\vartheta_j$, with $\vartheta_j=0$ corresponding to the north pole.  
Two monomers are fixed at  angles $\vartheta_\eta$ and $\vartheta_\beta$, while the third one (angle $\vartheta_\alpha$) is mobile within a harmonic trap. A wavepacket carrying a single electronic excitation approaching the Fano defect from the far left (yellow) may be reflected (red) or transmitted (green) by the latter. }
\label{geometry}
\end{figure}
However, incorporating electron-vibrational coupling into quantum transport studies is a challenging problem. It has been tackled using perturbative approaches for wires on the molecular or atomic scale \cite{montgomery2003electron,chen2005inelastic}, scattering theory \cite{sols1992scattering,ness1999quantum}, Green's function methods \cite{troisi2005modeling,jiang2005first,ness2006quantum}, master and quantum kinetic equations \cite{petrov2004spin}, reduced electron-density matrix approaches \cite{nitzan2001electron} or a semiclassical treatment of the motion and approaches based on non-equilibrium statistical physics \cite{hliwa2002tunnel}.

In the following, we formulate an alternative approach, which can exactly treat  quantum transport and a few vibrations in a multi-channel quantum scattering framework. We apply our theory to quantum transport on a long 1D chain or wire coupled to a control unit  (CU) that gives rise to Fano resonances in the transmission spectrum which is strongly modified by active vibrational modes  of the CU. To be specific we will cast our results in terms of a chain consisting of atoms or molecules. However, the results  more generally  apply to discrete chains, such as conjugated polymers \cite{kline2006morphology,hwang2011electronic} in the tight binding approximation, molecular wires \cite{ratner1998molecular,lehmann2004vibrational}, coupled quantum dots \cite{citrin1995coherent,kubota2011exciton,braakman2013long} with involvement of phonons or opto-mechanical arrays \cite{safavi:optomecharray,Schmidt:QMBoptomecharray}. Hence, reference  to a ``monomer'' in the subsequent text refers to any single site of the aforementioned transport systems.

Our article is organized as follows. To render our approach better comprehensible we first formalize our exemplary transport system in section \ref{model} including  a comprehensive description of the Fano-Anderson chain  in \sref{fanovib} and entanglement in \sref{entangle}. The new multichannel quantum scattering method is introduced in \sref{multichann}.  In \sref{ex_trans}  we apply the new method to investigate  excitation transport in the absence (\sref{im_trans}) and presence (\sref{mob_trans}) of vibrational motion. The results are summarized in \sref{concl}.

\section{Model and Methods} 
\label{model}
\subsection{Fano-Anderson chain with vibrating elements} 
\label{fanovib}
We consider a linear chain of $N$ monomers, which is in contact with a control unit containing three monomers confined on a circle of radius $R$, as sketched in \fref{geometry}. While many CUs are conceivable, our CU is small, convenient to design as we will discuss later, and most importantly,  vibrations of our control unit  can affect transport \emph{significantly}. We refer to the electronic state of the system with a single excitation on monomer or site $n$ as $\ket{\pi_n}$. For atomic or molecular systems this implies that only the $n$'th entity is excited, with all others in the ground-state. For other transport systems mentioned in the introduction, the state would imply a nearly empty lattice of sites, with a single particle filling the $n$'th site. The index $n$ fulfills $n \in \mathbb{Z} \cup \{\alpha, \beta, \eta\} $, with integer $n \in \mathbb{Z}$ indicating an excitation on the main chain and  $n \in \{\alpha, \beta, \eta\} $ on the CU, see \fref{geometry}.

Interaction between sites enables the excitation (e.g.~excited state or electron) to move along the chain, while conserving their total number, where we consider a single excitation only. For simplicity, we restrict ourselves to nearest neighbour interactions of strength $J$ in  the chain Hamiltonian
\begin{equation}
\hat{H}_{C} = J \sum_{n \in  \mathbb{Z}} \left( \ket{\pi_n}\bra{\pi_{n+1} }+ \ket{\pi_{n+1}}\bra{\pi_{n}} \right).
\label{ham_c}
\end{equation}
In addition to the electronic degrees of freedom discussed so far, we have to formalize
the vibrational degree of freedom in the CU which can couple to the electron dynamics. To this end
 we consider the angular coordinate $\vartheta_n$ of the control unit monomers $n \in \{\alpha, \beta, \eta\} $ to change according to the harmonic oscillator Hamiltonian 
\begin{subequations}
\begin{align}
\label{hamil_vib}
\hat{H}_{\rm vib} &= - \sum_{n\in \{\alpha, \beta, \eta\} }\frac{ \hbar^2\boldsymbol{\nabla}_{\vartheta_n}^2}{2I_n}  + \hat{V} (\boldsymbol{\vartheta}), \\
\hat{V} (\boldsymbol{\vartheta})&=\sum_{n\in  ( \alpha,\beta,\eta)} \frac{1}{2} M_n \omega^2 R^2 \Bigg(2\Big(1-\cos(\vartheta^{(0)}_{n} - \vartheta_n)\Big)\Bigg) \CR
&\approx \sum_{n\in  ( \alpha,\beta,\eta)} \frac{1}{2} M_n \omega^2 R^2 (\vartheta^{(0)}_{n} - \vartheta_n)^2,
\end{align}
\end{subequations}
with  the equilibrium positions $\vartheta^{(0)}_{n}$ and  the moment of inertia $I_n= M_nR^2$, while $M_n$ is the effective mass of monomer $n$ and $\boldsymbol{\vartheta}=[\vartheta_\alpha, \vartheta_\beta,  \vartheta_\gamma]^T$  is the vector of all angles. 

For simplicity, we consider the two monomers $\beta$ and $\eta$ so tightly confined in their respective harmonic potential with $\hbar\omega_{\beta}=\hbar\omega_{\eta}\gg J$,  that there is a negligible extension of their ground-state wavefunction. They then remain in the ground state for all energies considered and we can neglect their vibrational degrees of freedom. Only monomer $\alpha$ has a harmonic confinement $\hbar\omega\ll J$ such that several vibrational states $\ket{j}$ contribute. In that case, the Hamiltonian $\hat{H}_{\rm vib}'$ possesses a discrete energy spectrum given by 
$\hat{H}_{\rm vib}' \ket{\Phi_j}  = \mathcal{E}_j \ket{\Phi_j} =\hbar \omega(j+\frac{1}{2})  \ket{\Phi_j}$. The complete orthonormal basis set of the total Hilbert space is formed by the direct products 
$\ket{nj} = \ket{\pi_n}_{el}\otimes \ket{\Phi_j}_{\rm vib},  n \in \mathbb{Z} \cup \{\alpha,\beta,\eta\}$. 
In the joint Hilbert space, the total Hamiltonian of the system is given by
\begin{subequations}
\begin{equation}
\hat{H} = \hat{H}_{C}\otimes \sub{\id}{vib} + \hat{H}_{U} + \hat{H}_{UC}  + 
 \hat{H}_{\rm vib},
\label{ham}
\end{equation}
where $\sub{\id}{vib}$ is the identity in the vibrational space and the vibrational motion affects
the dynamics in the CU, described by $H_{U}$ and $H_{\rm vib}$, and also couples to the chain according to $H_{UC}$,
\begin{eqnarray}
\hat{H}_U &&= \sum_{n,n'\in(\alpha,\beta,\eta)} \sum_{jj'} F_{nn'}^{jj'} \ket{nj}\bra{n'j'},
\label{hamil_S_F}
\\
\hat{H}_{UC} &&= \sum_{jj'}G^{jj'} \left( \ket{0j}\bra{\alpha j'} +  \ket{\alpha j'}\bra{0j} \right),
\label{hamil_SC}
\\
 \hat{H}_{\rm vib} &&= \sum_{nj} \mathcal{E}_j  \ket{nj}\bra{nj}=\sub{\id}{el} \otimes \sum_j \mathcal{E}_j  \ket{j}\bra{j}.
\label{hamil_4}
\end{eqnarray}
where  $\sub{\id}{el}$ is the electronic identity.
The coupling matrix elements are
\begin{eqnarray}
F_{nn'}^{jj^{'}} &&= \int d {\boldsymbol{\vartheta}}\, \Phi_j^*({\boldsymbol{\vartheta}})F_{nn'}( {\boldsymbol{\vartheta}})\Phi_{j'}({\boldsymbol{\vartheta}}),
 \label{Fs_rep}
 \\
 G^{jj^{'}} &&=  \int d {\boldsymbol{\vartheta}}\, \Phi_j^*({\boldsymbol{\vartheta}})G( {\boldsymbol{\vartheta}})\Phi_{j'}({\boldsymbol{\vartheta}}),
 \label{s_rep}
\end{eqnarray}
\end{subequations}
where $F_{nn'}({\boldsymbol{\vartheta}})$ and $G({\boldsymbol{\vartheta}})$ are  $\boldsymbol{\vartheta}-$dependent hopping parameters. To keep our approach general enough to apply to diverse transport systems as discussed in the introduction, the basic formalism does not make reference to a specific model of interactions, except that  $\hat{H}_U$  and $\hat{H}_{UC}$  should depend on the spatial coordinates $\boldsymbol{\vartheta}$ of 
the control unit monomers. They therefore also depend on the two parameters $d$ and $R$ of the design, the distance from the center of the control unit 
 to the main chain, and the radius of the CU, respectively (see \fref{geometry} for the geometry). Note that through the dependence of the dynamics on $\boldsymbol{\vartheta}$,  the vibrational motion of the monomer in the ring will be coupled to the excitation transport on the chain. 
 For the presented numerical results, we use dipole-dipole interactions 
for which the hopping parameters scale with the inverse cubed distance between the sites, 
for details see \aref{app_evibcoupling}.

We are interested in the quantum transport of an excitation passing  the control unit on the chain from the left to the right. In the absence of the control unit  the transport can be described in terms of the eigenstates of \bref{ham_c}. They form an exciton band of energies $E_k = 2J\cos{k} ~\text{with}~ |E_k| \le 2 J$, where $k$ is the wavevector of the incoming excitation. We refer to this band as "continuous" in the following, implying the limit of infinitely many monomers in the chain, while for numerical calculations we employ of course a finite number of elements on the chain, $N=1000$. 
Subject to dispersion, the excitation can migrate across the chain.
To enable the CU to act as a scatterer and therefore to efficiently influence the transport, $J$ will be chosen of the order of all other energy scales in the system. 

This influence is mediated through the interactions between the continuous band of the linear chain and the three discrete eigenstates of the scatterer belonging to the CU and leads to modifications of the transmission characteristics by virtue of the  Fano resonance \cite{fano1935sullo,fano1961effects,miroshnichenko2010fano}. In the case of static interactions, i.e., without vibrations in $\boldsymbol{\vartheta}$, it is well understood how the control unit  affects transport on the chain \cite{miroshnichenko2010fano,miroshnichenko2005engineering,chakrabarti2007fano}. Specifically, for incoming wave energies that match an eigenenergy of the isolated static CU, transmission will be fully suppressed, as discussed later. The width of such a resonance dip in the transmission profile depends on the strength of the interaction $G$ between the main chain and the control unit in \bref{hamil_SC}.

\subsection{Quantum dynamics and entanglement} 
\label{entangle}

We now explore the fate of  an initial single excitation prepared on the far left side of the linear chain of monomers on its path across the chain. It must eventually impact the scattering region of the CU. 
In a time-dependent picture, we start from an initial state of the form,
\begin{equation}
\ket{\Psi(t=0)} = \ket{\sub{\psi}{ini}}\otimes \ket{j_\mathrm{in}}\,,
\label{in_qs}
\end{equation}
where  $\ket{j_\mathrm{in}}$ is the initial vibrational state, chosen as eigenstate of the vibrational Hamiltonian, and $\ket{\sub{\psi}{ini}}$ is the electronic excited state. We take it to be localized near a site $n_0{<}0$ with negligible amplitude on sites  $n{\ge}0$.  Specifically we choose $\langle  n | \sub{\psi}{ini}\rangle={\cal N}\exp{[-(n- n_0)^2/\sigma^2 
+i k_\mathrm{in} n]}$, where ${\cal N}$ is a normalisation factor and $k_\mathrm{in}$ the central incoming wave-number.
 The subsequent quantum dynamics of this wavepacket is governed by the time dependent Schr\"odinger equation (TDSE) $i\hbar\,{d}\Psi/{d}t=\hat{H}\Psi$
with
$\hat{H}$ from \bref{ham}. The state can be expanded as 
\begin{equation}
\ket{\Psi(t)}=\sum_{nj} \psi_{nj}(t) \ket{nj}.
\label{entangled_state}
\end{equation}

\begin{figure}[h]
\includegraphics[width=\columnwidth]{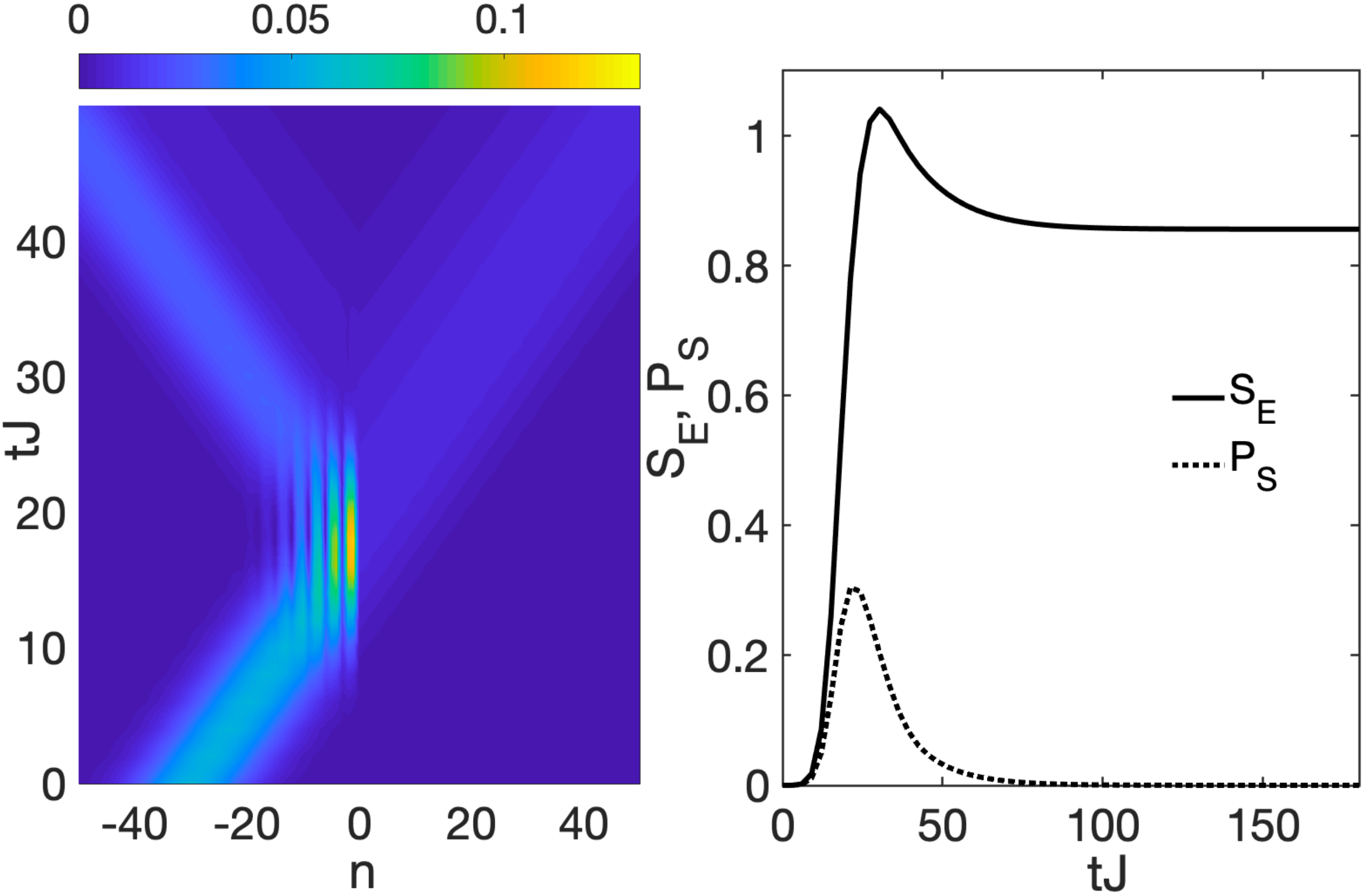}
\caption{ (a) Evolution of the excitation probability $p_n=\sum_j |\psi_{nj}(t)|^2$on the main chain, 
 allowing vibrations of site $\alpha$.  Populations on the CU are not shown. The vibration frequency is $\hbar\omega = 0.01J$ and other parameters are given in footnote \cite{par_1} for a chain with $1000$  monomers.  An excitation initially localized near $n=-30$ moves towards the right with energy $E=1J$ and reflects back from the scattering region. The x-axis indicates the position on the chain in terms of monomer indices $n$. (b) The von Neumann entropy $S_E$ \eqref{ent_eq} as a function of time. 
The total excitation probability $P_U=\sum_{n\in\{\alpha,\beta,\eta\}j} 
|\psi_{nj}(t)|^2$ on the scattering control unit $U$ is also shown.\label{entropy}}
\end{figure}

An example of such a scattering process is shown in \fref{entropy}(a), where the time-dependence of the population on the sites $n$ of the chain is displayed. Populations of CU monomers are not shown.
One sees that the incoming wavepacket moves with constant velocity.
Close to the time $t_{c}$ when the center of the wavepacket reaches the  site $n=0$, the scattering site which is closest to the CU, an interference pattern appears. After the collision for $t>t_{c}$, one can clearly see the transmitted and reflected wavepacket.

For times $t<t_{c}$, before the excitation reaches the CU, the vibrational degrees of freedom of the CU  remain in their initial state $\ket{j_\mathrm{in}}$ and  the dynamics of the system is  exclusively governed by $\hat{H}_C$. As the excitation hits the scattering region, the other terms in the Hamiltonian become important and vibronic quantum dynamics ensues for a finite time interval, until the excitation  completely leaves the scattering 
region. The post-collision dynamics is again governed by $\hat{H}_C$. The 
final outcome of the scattering event is the splitting of the excitation probability
  into transmitted (right moving) and reflected (left moving) parts and a possible change of the internal vibrational state of the scatterer into some general superposition of vibrational eigenstates. Each vibrational state contributes to the outcome of the scattering 
process leaving multiple outgoing channels for the scattering. 

An immediate consequence of the dependence of the scattering outcome on vibrational states of the CU is the
creation of entanglement between the electronic and vibrational states during the scattering process. Entanglement can be quantified by the von-Neumann entropy
\begin{equation}
S_E= - \mbox{tr}(\sub{\rho}{el} \ln \sub{\rho}{el})\,,
\label{ent_eq}
\end{equation}
where tr denotes the trace and $\sub{\rho}{el}$ the reduced electronic density matrix obtained by tracing out the vibrational degrees of freedom of the system \cite{bengtsson_zyczkowski_2006}, i.e.~$\sub{\rho}{el}=\sum_j \bra{j} \hat{\rho}\ket{j}$, with $\hat{\rho}=\ket{\Psi(t)}\bra{\Psi(t)}$ and $\ket{\Psi(t)}$ in \eref{entangled_state}. This entropy is zero if the electronic and vibrational degrees of freedom are separable
and equal to $\ln D$ for a maximally entangled state, if the reduced Hilbert-space is $D$-dimensional. 
We see in \fref{entropy}(b) that
before the wavepacket hits the scattering region, the electronic and vibrational states are not entangled and the entropy is zero. As expected  the entropy increases close to the impact time $t_{c}$, indicating the development of entanglement between  electronic and vibrational states.  This entanglement
persists for large times, even after the excitation on the CU (dashed line in  \fref{entropy}(b)) has dropped to zero again.

Conceptually, this means that even at a time $t=100/J$ when the exciton wavepacket 
has mostly returned to the main chain it remains intricately linked with the scatterer through entanglement with vibrational states.
The entanglement  leads to practical difficulties using conventional 
approaches such as the transfer matrix method (TMM). The electronic and vibrational part cannot be treated separately. Transmission, however,  depends on the vibrational state giving rise to
multiple channels. The incorporation of 
those channels for scattering leads to a cumbersome set of nonlinear 
equations in the TMM.

Therefore,  we develop in the following a multichannel quantum scattering method (QSM) that can handle the entanglement discussed above, as well as the 
effect of internal  vibrations of the CU on quantum transport through the chain.

\section{Multi-channel quantum scattering method (QSM)}
\label{multichann}

While we will use solutions of the TDSE for verification, we switch now to a time-independent framework, in which scattering processes are usually more easily understood 
based on stationary scattering states. Consequently, we seek a solution of the eigenvalue problem
\begin{equation}
\hat{H}\ket{\Psi}= E\ket{\Psi}
\label{se_stat}
\end{equation}
with 
\begin{equation}
\label{eq:Psi}
\ket{\Psi}=\sum_{nj}\psi_{nj} \ket{nj}\,,
\end{equation}
in analogy to the expansion of the time-dependent state \bref{entangled_state}.
We intend to solve \bref{se_stat} for the case of an exciton wave coming 
in from the left with momentum $k_\mathrm{in}{>}0$ and the CU in 
the specific state $\ket{j_\mathrm{in}}$ with  vibrational quantum number $j_\mathrm{in}$. (More complex initial vibrational states could be generated from these solutions by 
superposition.) 
 The desired eigenstate $\ket{\Psi}$ is subject to the boundary conditions 
\begin{subequations}
\begin{align}
\psi_{nj} &= \delta_{jj_\mathrm{in}}e^{+ik_\mathrm{in}n}+\mathfrak{R}_j(k_\mathrm{in}) e^{-ik_{j}n}, \:\:\: n<0
\label{b_cond1}
\end{align}
for $n$ on the left side of the chain and  analogously for
 $n$ on the right side of the chain
\begin{align}
\psi_{nj} &= \mathfrak{T}_j(k_\mathrm{in}) e^{+ik_{j}n}, \:\:\: n>0.
\label{b_cond2}
\end{align}
\end{subequations}
Here, $\mathfrak{R}_j(k_\mathrm{in})$ and $\mathfrak{T}_j(k_\mathrm{in})$ 
are complex reflection and transmission amplitudes containing all information about the scattering outcome, and the momentum $k_j > 0$ of the $j^{th}$ channel is fixed by energy conservation $E =  \mathcal{E}_{j_\mathrm{in}} + E_{k_\mathrm{in}}=  \mathcal{E}_{j} + E_{k_{j}} $ for the exciton band energies  $E_k$ and the vibrational energies $\mathcal{E}_{j}$, as defined earlier.
The probability of transmission $T_j(k_\mathrm{in})$  is given by the ratio between the transmitted  and  incoming flux in  channel $j$,
\begin{subequations}
\begin{eqnarray}
T_j(k_\mathrm{in}) = | \mathfrak{T}_j(k_\mathrm{in})| \frac{\sin{k_j}}{\sin{k_\mathrm{in}}}.
 \label{trans}
\end{eqnarray}
Similarly, the reflection coefficient $R_j(k_\mathrm{in})$  is given by the ratio between reflected  and  incoming flux in  channel $j$,
\begin{eqnarray}
R_j(k_\mathrm{in}) = | \mathfrak{R}_j(k_\mathrm{in})| \frac{\sin{k_j}}{\sin{k_\mathrm{in}}}.
 \label{refl}
\end{eqnarray}
Conservation of probability implies 

\begin{eqnarray}
\sum_j \left[ T_j(k_\mathrm{in}) + R_j(k_\mathrm{in})  \right] = 1,
 \label{conserve}
\end{eqnarray}
\end{subequations}
and can serve as a consistency check of the final results.

In order to construct the desired stationary scattering state $\ket{\Psi}$, one needs to substitute the boundary conditions \bref{b_cond1} and \bref{b_cond2} back into the time independent Schr\"odinger equation \bref{se_stat} and solve for $\mathfrak{R}_j$ and $\mathfrak{T}_j$.  Exciton-phonon coupling links all vibrational components of the wave function, rendering the  equation hard to solve for many vibrational levels of the scatterer.

Instead of tackling this problem directly, we  employ a well established trick \cite{sukhanov2014remote,kuzovova2018solution} based on backward propagation.
We convert the original \eref{se_stat} into a collection of auxiliary eigenproblems
\begin{subequations}
\begin{eqnarray}
\hat{H} \ket{\Psi^{(j_0)}} = E \ket{\Psi^{(j_0)}},
\label{aux_latt_eqn}
\end{eqnarray}
subject to new boundary conditions
\begin{align}
\psi^{(j_0)}_{nj} &= A_j^{j_0}  e^{+ik_{j}n} + (\delta_{jj_0} - A_j^{j_0}) e^{-ik_{j}n}  \hskip0.2cm (n\le 0)\label{aux_eigfn} \\
\psi^{(j_0)}_{nj} &= \delta_{jj_0} e^{+ik_{j_0}n} \hskip3.2cm (n\ge 0)  \,.
\label{aux_eigfn01}
\end{align}
\end{subequations}
Here, the vibrational index $j_0$ is fixed but arbitrary, not related to $j_\mathrm{in}$. We thus specify the vibrational quantum number
for the outgoing part of the wave, instead of the incoming one.
Using these auxiliary solutions, one can then form a linear combination
that solves the original problem \bref{se_stat} as
\begin{eqnarray}
\psi_{nj} = \sum_{j_0} C_{j_0}\psi^{(j_0)}_{nj}.
\label{aux_eigfn1}
\end{eqnarray}
The coefficients $A_j^{j_0}$ in \esref{aux_eigfn}{aux_eigfn01} are  determined such that \bref{aux_latt_eqn} can be solved, as discussed in  \aref{app_reformulation}. Demanding that the linear combination \bref{aux_eigfn1} satisfies the original boundary conditions \esref{b_cond1}{b_cond2}, we obtain the following system of equations for the coefficients $C_{j_0}$:
\begin{subequations}
\begin{align}
 \sum_{j_0} A_j^{j_0} C_{j_0} &= \delta_{jj_\mathrm{in}},\\
C_j &= \mathfrak{T}_j(k_\mathrm{in}),\\
\sum_{j_0} (\delta_{jj_0}-A_j^{j_0})C_{j_0}&= \mathfrak{R}_j(k_\mathrm{in}).
 \label{aux_eigfn2}
\end{align}
\end{subequations}
Since the matrix $[A_j^{j_0}]$ is always regular, we can find its inverse 
$Q_{j_0}^{j}$ and then the coefficients
$C_{j_0} = Q_{j_0}^{j_\mathrm{in}}$ of the expansion \bref{aux_eigfn1}.
After the $C_{j_0}$ are determined, the reflection and transmission amplitudes $\mathfrak{R}_j(k_\mathrm{in})$ and $\mathfrak{T}_j(k_\mathrm{in})$ 
 simply follow.  The main advantage of this method over directly solving the eigenvalue problem \eref{se_stat} with the boundary conditions \eref{b_cond1} and \eref{b_cond2} is that the choice of the boundary conditions \eref{aux_eigfn01} eases the burden of obtaining the probability amplitude on the scatterers, which are essential to solve the problem (\aref{app_reformulation}).

The method can be implemented for any configuration of the scatterer without any restrictions on the number of monomers in the scatterer. It can even 
 be modified for cases in which the scatterer interacts with several monomers of the main chain. The contribution to the transmission profile from each channel can be explicitly obtained and an estimate of the final quantum state of the scatterer can also be deduced. 

Next, we apply our new QSM approach to the transport system with a CU containing an active vibrational degree
of freedom.

\section{Excitation Transport} 
\label{ex_trans}
\subsection{Static Monomers}
\label{im_trans}

Excitation transport in a linear chain of monomers interacting with a static CU, i.e., in the absence of any vibrations, was explored in the past using a transfer matrix method (TMM) to obtain the transmission and reflection coefficients \cite{miroshnichenko2010fano}, see also \aref{tmm}. Transport was found to be highly sensitive to the resonance properties of the CU, which acts as a defect \cite{miroshnichenko2005engineering}. Due to the interference between the continuum energies of the main chain and the discrete energies of the control unit, the system exhibits Fano resonances.  


We consider the configuration where the CU monomers form a static equilateral triangle,
 corresponding to angles  $\boldsymbol{\vartheta} =  [0,2\pi/3, 4\pi/3]^T$ 
, 
 with a two-fold goal: Firstly,  to illustrate the features above, and secondly to benchmark the QSM formulated in \sref{multichann} by comparison with the TMM and with solutions of the TDSE. 
It was shown in \cite{tong1999wave,miroshnichenko2005nonlinear,miroshnichenko2010fano} that for an incoming energy $E$ the transmission amplitude  is
\begin{subequations}
\label{transmission_profile_TMM}
\begin{eqnarray}
T (E) &=& \frac{4J^2 - E^2}{4J^2 - E^2+\sub{V}{eff}(E)^2}\\
\sub{V}{\rm eff}(E)  &=& \frac{D_2}{D_3}G^2\,
\end{eqnarray}
\end{subequations}
with the energy dependent effective scattering potential  $V_{\rm eff}(E)$ from the CU. The latter is determined
 in the chosen configuration through $D_i = \mbox{det}(E \mathbb{1} - H_i)$, where $H_3$ and $H_2$  are the dipole-dipole Hamiltonian of the CU and  of the CU without the entrance site $\alpha$ respectively, as derived in \aref{tmm}. One sees from \bref{transmission_profile_TMM}, that transmission is completely suppressed if $V_{\rm eff}$ tends to infinity,
 which happens if $D_{3}=0$, i.e., whenever the energy $E$ coincides with an eigen energy of $H_{3}$.
 On the other hand, if the energy $E$ matches an eigen energy of $H_{2}$ we have $D_{2}=0$ implying that $V_{\rm eff}=0$, and therefore the transmission is maximal, $T=1$. Furthermore, for any finite value of $V_{\rm eff}$, transmission is fully suppressed at $E = \pm 2J$.

\begin{figure}[h]
\includegraphics[width=0.9\columnwidth]{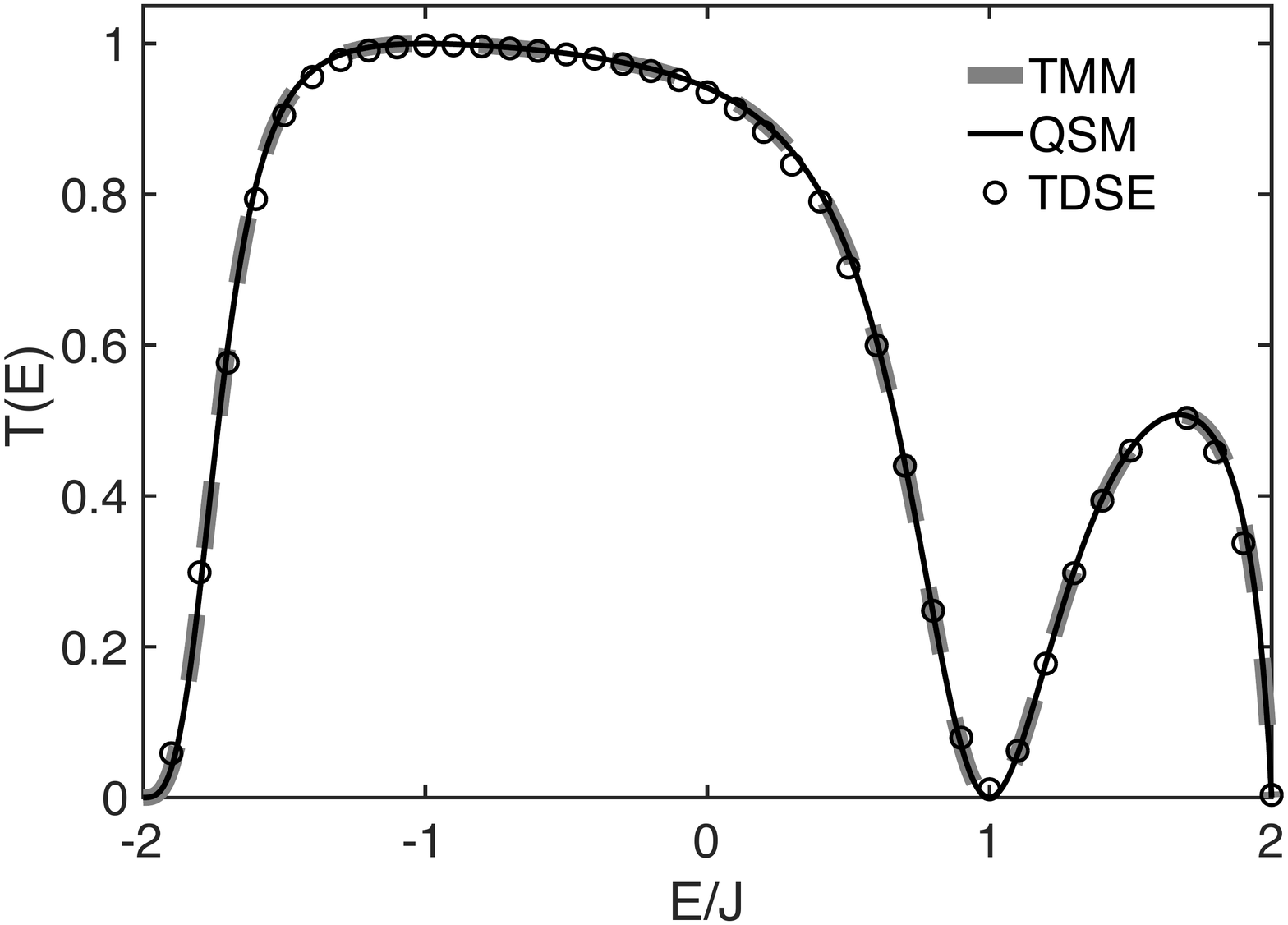}
\caption{Total transmission through the chain as a function of energy for fixed monomers
and the control unit forming an equilateral triangle with further parameters given in  \cite{par_1}.
Results obtained with the quantum scattering method (QSM) (thin solid line), with a time dependent wavepacket propagation ($\circ$) and the transfer matrix method (TMM) (thick dashed line) are shown. \label{trans_/static}}
\end{figure}

These properties can be directly found in the transmission profile  shown in \fref{trans_/static}. Firstly, one can appreciate that results from the TMM and calculated with the more complex QSM formalism in 
\sref{multichann} as well as TDSE solutions agree well. The CU $\hat{H}_U=H_{3}$ has two degenerate eigenenergies at $E=J$ and a third one at $E=-2J$, since we choose $F=F_{\alpha\beta}=F_{\alpha\eta}=F_{\beta\eta}=J$.
In contrast the reduced control unit without the entrance site, $\hat{H}'_U=H_{2}$ has two eigenenergies, $E_{\pm}=\pm $J. Hence we expect transmission extrema at $E=\pm 2J$ and $E=\pm J$, more precisely zero transmission at energies $\pm 2J$ and $J$ and full transmission at $ E=-J$, which is in accordance with \fref{trans_/static}.
The complete suppression of transport at $J=1$ has an asymmetric profile, characteristic for a Fano resonance. Its width depends on the interaction strength between the main chain and the CU determined by the relative position of the CU with respect to the chain. It is clearly sensitive to details of the CU since  $E=J$ is an eigenenergy for both, $H_{2}$ and $H_{3}$. However, the influence of the latter dominates, as its eigenstate is doubly degenerate.

The QSM has been developed to study the transport properties for the vibrating CU with a motional degree of freedom. To compare it with the static CU case discussed in this section, we only allow the ground-state $j=0$ in all sums of \sref{multichann}, effectively freezing the motional degree of freedom. We also call this scenario ``immobile''.
As can be seen in \fref{trans_/static}, the immobile QSM is equivalent to a static calculation using the TMM.
The  QSM for an immobile CU and the TMM require comparable computational effort.
 While the TMM takes into account an effective scattering potential, QSM exactly determines the state of the scatterer and  provides the transmission and reflection coefficients through proper boundary conditions in the Schr\"odinger equation.

For final verification, we also solve the time-dependent Schr\"odinger equation (TDSE) with the complete Hamiltonian \bref{ham}. To obtain the transmission profile,  we take the Gaussian wavepacket of electronic states $\ket{\sub{\psi}{ini}}$ introduced in \sref{entangle} 
on the far left of the linear chain as the initial condition for the incoming excitation. The integrated  transmitted probability after the excitation has left the scattering region provides the transmission coefficient.
These numerical solutions are obtained using XMDS \cite{dennis2013xmds2,xmds2}.
We see in \fref{trans_/static} that the transmission coefficient obtained from the TDSE matches well with the QSM and TMM results, with minor deviations caused by the finite energy-width of the Gaussian wavepacket.
This enables us to use the TDSE solutions to verify the results obtained from the QSM also in the case of vibrations in the CU, which is our final goal and discussed in the next section.

%
\subsection{Fano Resonances with Vibrations}
\label{mob_trans}

To see how electronic--vibrational coupling affects the quantum transport, we now mobilize monomer $\alpha$
of the CU such that it can execute small harmonic vibrations with frequency $\omega$ on the circle around its equilibrium position. Since $\hbar\omega\sim J$, several vibrational
levels with energy $\hbar\omega(j+1/2)$ can be excited from the vibrational ground state $j_{\rm in} = 0$ by the incoming electronic wavepacket
with energies of the exciton band in the range $[-2J,2J]$. Constraining all but one monomer is for simplicity only, our methods can be generalized to include vibrations of all control unit monomers.

\begin{figure}
\includegraphics[width=0.8\columnwidth]{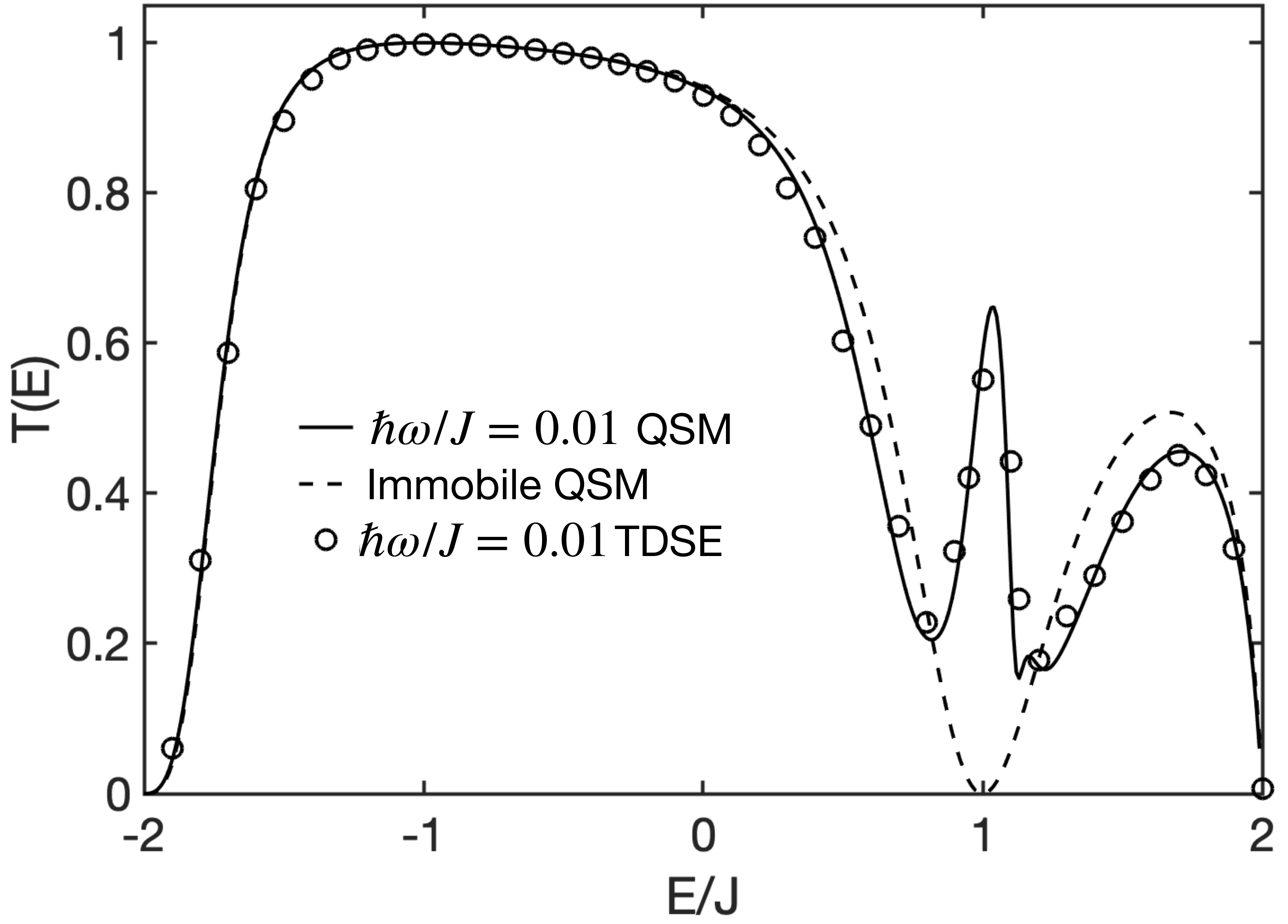}
\caption{Comparison of transmission profile for vibrating and immobile monomer $\alpha$ in the control unit. The dashed curve shows the transmission profile obtained from our multi-channel quantum scattering method (QSM) when the vibrational state is constrained to the ground-state. The solid curve shows the transmission profile obtained from the QSM for the vibrating monomer, with harmonic frequency $\hbar\omega/J = 0.01$ and other parameters as in \cite{par_1}.  Circles show the transmission profile obtained from a time dependent wavepacket calculation for verification. For the QSM, here and elsewhere, we use $50$ vibrational states unless otherwise indicated,  with results unchanged for higher numbers. Here and in the subsequent plots,  monomer $\alpha$ is assumed to be initially in the vibrational ground state $j_\mathrm{in} = 0$.}
\label{trans_vib}
\end{figure}

In \fref{trans_vib}, we compare the transmission profile obtained from the QSM for the immobile and mobile scatterer. 
Allowing vibrations (here with a small frequency $\hbar\omega=0.01 J$) clearly modifies the Fano profile
most significantly close to the Fano resonance dip at $E=J$, where we see finite transmission instead
of full reflection  in the immobile case. 
 In contrast, vibrations leave the other regions of the transmission spectrum largely unaffected. In particular the perfect transmission at $E = -J$ and reflection at $E=\pm 2J$ persist. This may be understood realizing  that these characteristics are due to the reduced
CU  $\hat H'_{U}$  and the chain just in the presence of a CU, respectively. The conditions for both of these elements remain the same if $\alpha$ is mobilized. 

Empirically, a CU with three sites is the minimal configuration to easily achieve the desired large influence of vibrations, i.e., significant broadband transmission in the presence of vibrations for incoming energies that are opaque in the static case. This can be understood through the fragility of the quotient $D_{2}/D_{3}$ in the static case \eref{transmission_profile_TMM}, for the case where numerator and denominator both tend to zero.  This control unit thus gives rise to a \emph{qualitative impact} of vibrations on transport. Strictly speaking, a single vibrating monomer in the control unit is sufficient to influence transport on the chain, although, to a much smaller extent
and only for a very narrow regime of parameters which required excessive parameter  tuning to find.
Moreover, for a single  monomer in the CU, both the coupling between CU and main chain as well as the coupling strength to higher vibrational excitations  depend on a single physical parameter, the interaction strength between CU and main chain. For more monomers in the CU, the coupling strength to higher vibrational excitations additionally depends on intra-CU interactions, allowing the two crucial quantities to be independently tuned.

For verification of the QSM results, we compare them to those obtained with the TDSE. In contrast to the previous section, we explicitly include the vibrational dynamics of monomer $\alpha$ in the simulation, which is initialised in the vibrational ground state. After the scattering event,
the transmission coefficient  shown in \fref{trans_vib} is obtained as discussed before, summing over all vibrational channels. The TDSE quantitatively confirms the transmission profile, in particular the
spectacular switch of the suppression  around $E=J$ to a local maximum of transmission due to an excitable  CU.

\begin{figure}
\includegraphics[width=\columnwidth]{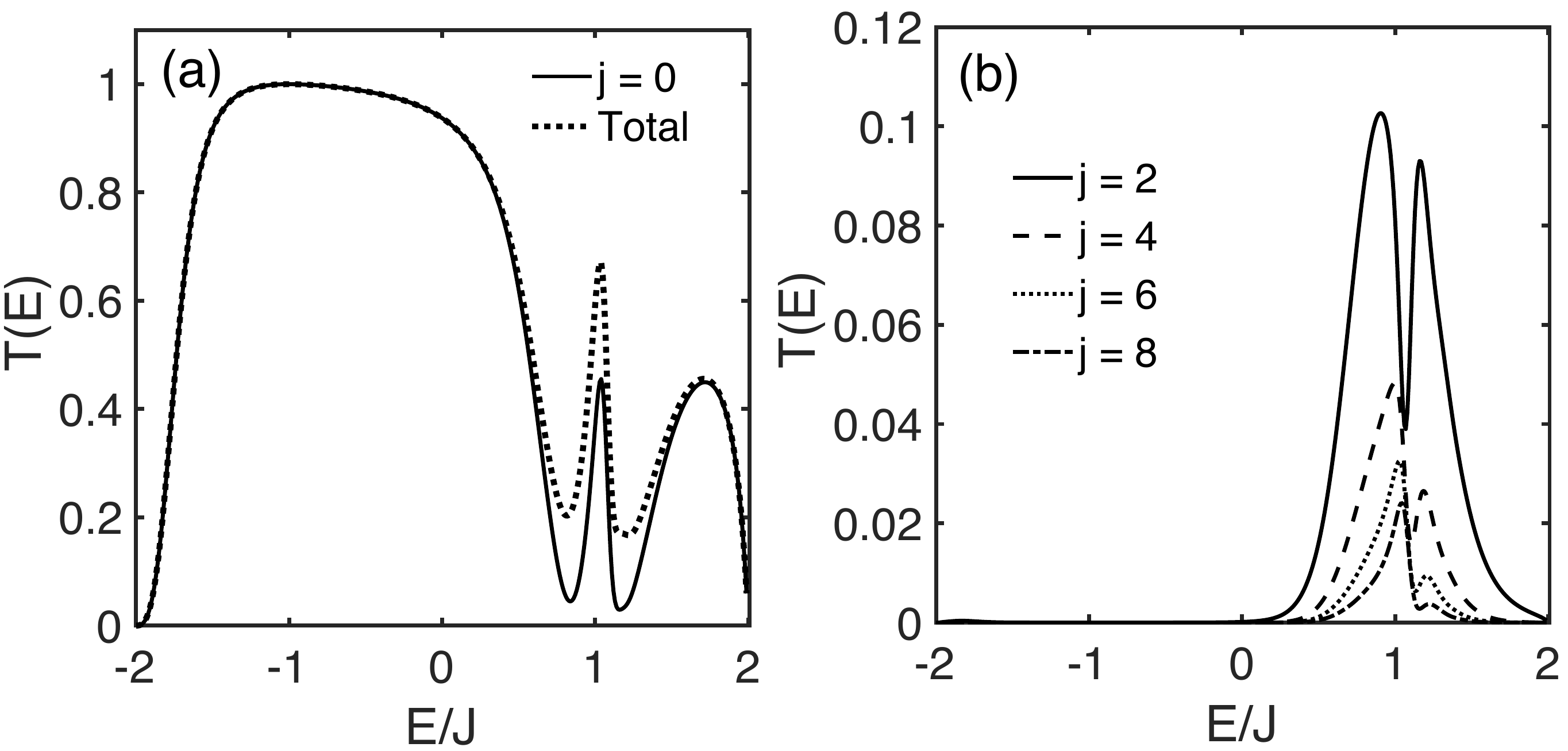}
\caption{Contributions of different vibrational channels: (a) Total transmission profile (dotted line), compared to the elastic channel, i.e., for $j_{\rm in}=j_{\rm final}=0$ (solid line). (b) Contribution from inelastic channels with $j_{\rm final}>0$. The harmonic frequency for monomer $\alpha$ is $\hbar\omega/J = 
0.01$ and other parameters as in \cite{par_1}. }
\label{trans_vib_chan}
\end{figure}
%

\subsubsection{Distinguishing vibrational channels}

The QSM allows us to separately quantify the contribution of each vibrational channel $j$ as done in \fref{trans_vib_chan}, where the individual transmission probabilities $T_j$ are shown together with the total transmission $T=\sum_jT_j$. 
Partial transmission $T_j$ means that the vibrating
monomer remains in vibrational state $\ket{j}$ \emph{after} the excitation has passed the control unit region, and hence can no longer affect the vibrational state.
One sees from \frefp{trans_vib_chan}{a} that the elastic channel without lasting energy exchange between main chain and CU dominates transmission apart from energies close to the resonance at $E=J$. This remains true
even if the initial vibrational state for the monomer is not the ground-state, as in \fref{trans_vib_chan}, but rather an excited state. We see that only even $j$ contribute here, due to the mirror symmetry of the setup in \fref{sketch}. Without this symmetry, also odd $j$ would contribute.

The transmission profile associated with the $\ket{j_\mathrm{in}}$ channel is similar to the static case, except in the vicinity of $E=J$ where a non-zero transmission is observed. Other channels contribute quite significantly to transmission in the dip region, where a small finite transmission contribution is found, which decreases with the inelasticity, i.e., with increasing vibrational energy of the channel, see \fref{trans_vib_chan}{b}. The sum of total 
transmission and reflection over all channels is unity for the entire energy region, confirming the consistency of the method.

\subsubsection{Vibrational resonances}
\label{vibres}

So far we have focused on small vibrational frequencies $\hbar\omega\ll J$.
 High vibrational frequencies $\hbar\omega>2J$ lead to quantized vibrational states outside the exciton band which are weakly coupled for our parameters.
%
\begin{figure}
\includegraphics[width=\columnwidth]{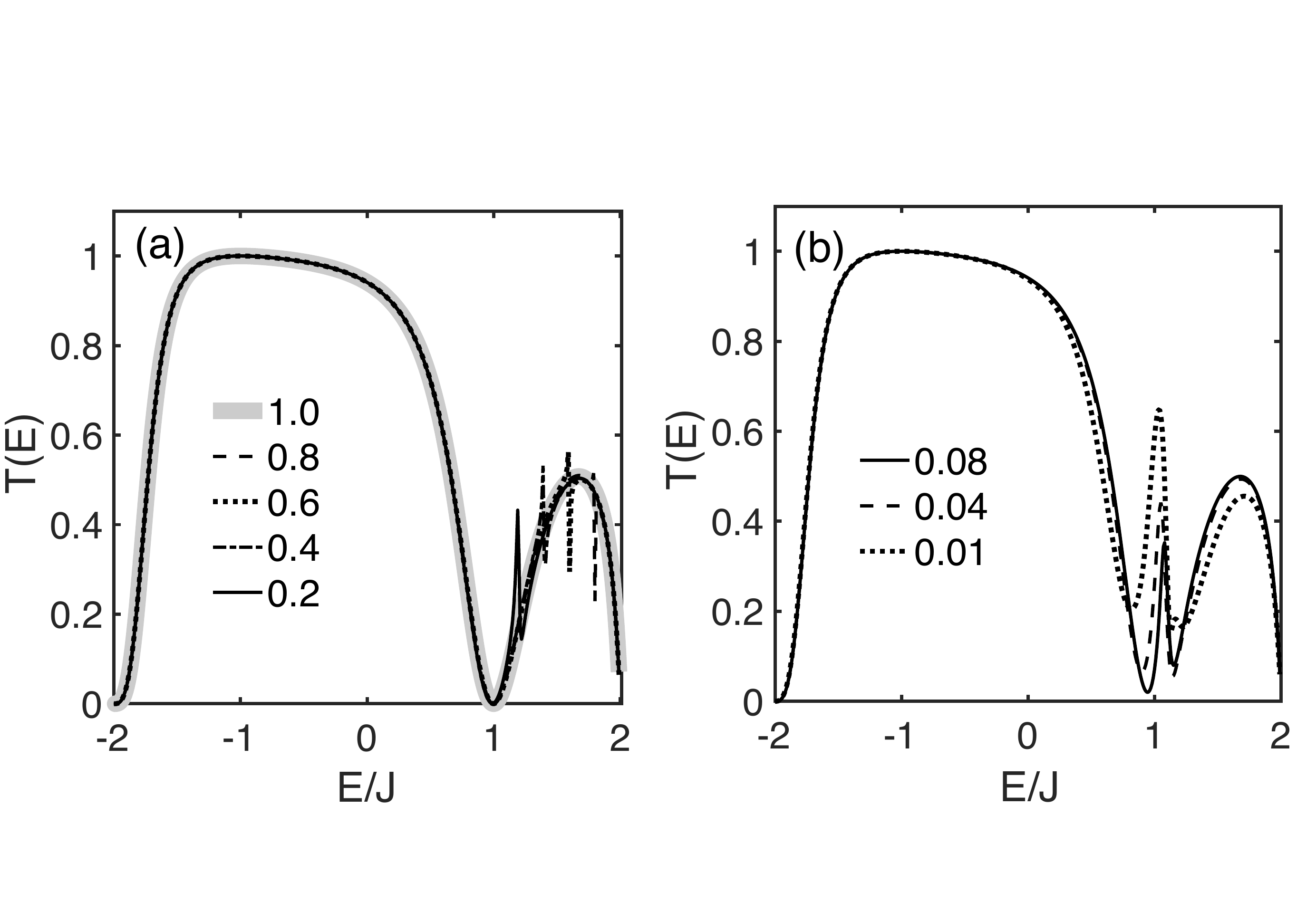}
\caption{ Transmission as a function of energy for different vibration frequencies of the ring monomer with parameters given in \cite{par_1}. The legends indicate the vibrational energy $\hbar\omega/J$. (a)  Frequencies $\hbar\omega/J\geq 0.2$. (b) Frequencies $\hbar\omega/J< 0.2$.
\label{trans_vib_detail}}
\end{figure}
The transmission profile for the excitation transport is shown in \fref{trans_vib_detail} for several vibrational frequencies $\omega$. As expected, if  $E=J+\hbar \omega$ falls outside the exciton bandwidth, here for $\hbar \omega \ge J$, the transmission profile is not affected by the vibration, see thick gray curve in \frefp{trans_vib_detail}{a}. For lower frequencies a clear feature appears at $E=J+\hbar \omega$, in the form of a narrow peak and dip, superimposing the already existing broad dip centred at $E=J$. The characteristic profile seen again heralds a Fano resonance that now involves the vibration of the monomer in addition to electronic degrees of freedom. As $\omega$ is further reduced, the resonance peaks
move towards $E=J$ and broaden giving rise to transmission instead of reflection at $J=1$ as discussed before, see \frefp{trans_vib}{b}. 

\begin{figure}[htb]
\includegraphics[width=0.9\columnwidth]{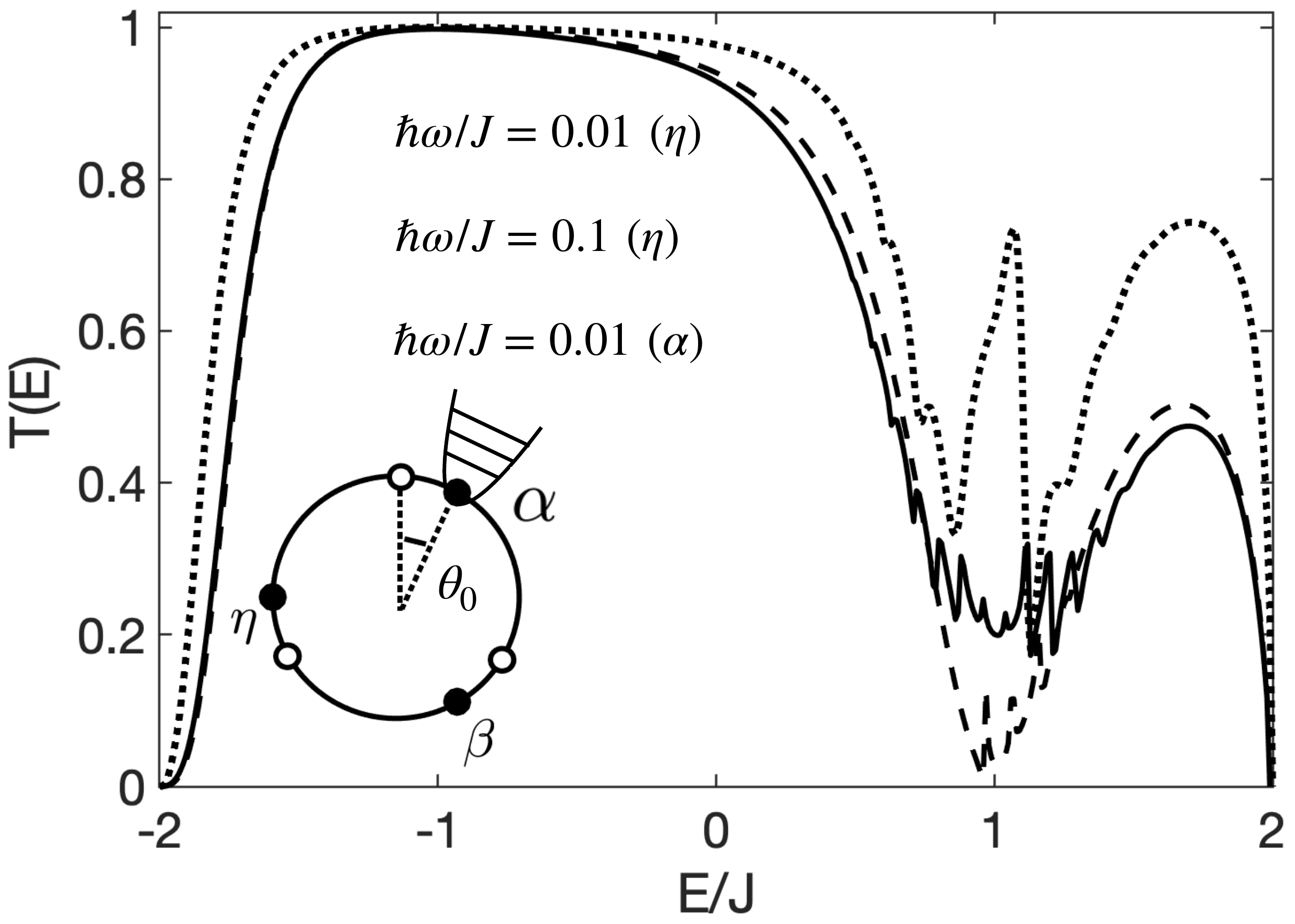}
\caption{Transmission as a function of energy for two harmonic frequencies  $\omega_{\eta}$ of the ring monomer $\eta$ with  parameters given in \cite{par_1} and the two other monomers of the CU kept immobile. 
Transmission spectrum (dotted) for an asymmetric CU configuration where only monomer $\alpha$ is mobile and
the CU has been rotated
 by $\theta_0=\pi/6$ (see inset) relative to the main chain.} 
\label{trans_eta}
\end{figure}
%

\subsubsection{More vibrations and transmission switching}

While we have only considered vibrations on monomer $\alpha$ so far, let us briefly inspect what happens if monomer $\eta$ is mobile instead ($\eta$ and $\beta$ are equivalent by symmetry). For  $\hbar\omega_{\eta}/J\ge 0.1$ (dashed line in  \fref{trans_eta}) the transmission profile is qualitatively similar to the case of a static CU, while multiple resonance kinks appear for the smaller frequency $\hbar\omega/J = 0.01$ (solid line in  \fref{trans_eta}) turning the transmission dip into a region of finite transmission. 

Note that monomers $\alpha$ and $\eta$ play a very different role for our transport system
which is most easily seen in the static case from our discussion of \eref{transmission_profile_TMM}, where the eigenenergies
of the CU with $\alpha$, $D_{3}$ and the eigenergies of the reduced CU $H_{2}$ (including $\eta$ but excluding $\alpha$) enter the effective potential $V_{\rm eff}$  as a factor $D_{2}/D_{3}$.
The difference of the roles can be blurred with an asymmetric CU configuration which is achieved by simply rotating the entire CU with an angle  $\theta_{0}$ as shown in the inset of \fref{trans_eta}. Indeed, now the transmission for 
a mobile monomer $\alpha$ keeping the other two monomers in the CU immobile,  shows additional resonance features similar to a mobile $\eta$ before, while retaining the overall characteristics of mobile $\alpha$ from the symmetric case with a sizeable transmission at $E=J$, see dotted line in \fref{trans_eta}.

Importantly, in either case, with vibrations on $\eta$ or $\alpha$, we find that a spectral region of perfect reflection can be turned into one with significant finite transmission. This signals a complete qualitative change 
of resonant scattering  through  motion of the CU monomers. A general understanding of this behavior is provided by  non-adiabatic transitions between chain states discussed in \cite{ramachandran:nonadiabtransp} within an appropriate  time-dependent framework.

\section{Conclusions and Outlook}
\label{concl}
To describe the effect of vibrating sites in discrete transport systems we have  developed  a multichannel quantum scattering method (QSM)  which allows us to determine transmission and reflection coefficients in a time-independent framework, despite the strong coupling of excitation transport to vibration and creation of electronic-vibrational entanglement. We have verified the results and the QSM developed,  by extensive comparison with time-dependent wave-packet calculations using the TDSE.
Since  a larger number of vibrating sites can be included, the method is applicable in a general context
of transport on a discrete chain of sites with coupling to vibration or inter-site motion, ranging from conjugated polymers and molecular wires and coupled quantum dots with involvement of phonons to opto-mechanical arrays. 

Using this new method, we have explored how Fano-resonances in quantum transmission 
on a static chain of discrete sites (monomers) including a control unit (scatterer) are modified if 
the monomers are allowed to vibrate.
 This setup constitutes a Fano-Anderson chain with mobile scatterers. 
It gives rise to rich features, including the reversal of the scattering effect: 
Mobile scatterers can lead to  significant transmission at 
 incoming wave energies with full reflection in the static case.
 The qualitative difference of the transmission characteristics close to a Fano 
resonance with significant transmission upon different kinds of monomer mobilization in the control unit
 suggests possible applications in nanoscopic switching and sensing. 
 
\acknowledgments
We thank Milan \v{S}indelka for input to the initial stage of this project and the Max-Planck society for financial support under the MPG-IISER partner group program.  Also the support and the resources provided by Centre for Development of Advanced Computing (C-DAC) and the National Supercomputing Mission (NSM), Government of India are gratefully acknowledged. A.E.~acknowledges support from the DFG via a Heisenberg fellowship (Grant No EI 872/5-1). A. S. acknowledges financial support from SERB via the grant (File No.: CRG/2019/003447)  and from DST via the DST-INSPIRE Faculty Award (DST/INSPIRE/04/2014/002461).

\bibliography{ref}

\begin{thebibliography}{56}
\bibitem[{par()}]{par_1}
\bibinfo{note}{In all calculations, the radius of the circle is $R = 0.5774 \:
  \delta$, $d= 1.5774 \: \delta$, with distances defined in \fref{geometry}. In
  the present study, ${\vartheta}_{\beta\alpha}={\vartheta}_{\eta\beta}=2\pi/3$
  and monomer onsite energies in the ring are $E_n= 0 J$. This configuration
  make sure that there are eigenvalues of $\sub{\hat{H}}{el}^S$ within the
  exciton energy band of the main chain $|E|<2J$, which is essential for
  providing a Fano resonance. The mass $M$ is selected such that if $\delta = 1
  ~\text{a.u.}$, $M=11000 ~\text{a.u.}$. This mass corresponds to a Lithium
  atom, for a concrete example where the chain is assembled with
  dipole-dipole-interacting ultracold Rydberg alkali atoms.}
\end{thebibliography}
\appendix   

\section{Vibrational channel expansion coefficients}
\label{app_reformulation}
%
Here, we describe the steps to obtain the coefficients $A_j^{j_0}$ in \esref{aux_eigfn}{aux_eigfn01}. To solve the eigenvalue problem \bref{aux_latt_eqn}, let us first look at the action of the Hamiltonian $\hat{H}$ on the state $\ket{\Psi}$ defined in \eref{eq:Psi}.
Noting that  $\langle nj | \Psi\rangle=\psi_{nj}$, we find
\begin{equation}
\begin{split}
\bra{nj}\hat{H}\ket{\Psi} &= \Big( 1 - (\delta_{n\alpha}+\delta_{n\beta}+\delta_{n\eta})\Big)\Big(J \psi_{n+1,j}+\\&+J\psi_{n-1,j}\Big)+ \mathcal{E}_j \psi_{nj}\\
&+ (\delta_{n\alpha}+\delta_{n\beta}+\delta_{n\eta})   \sum_{n'\in(\alpha,\beta,\eta)}^{n'\ne n} \sum_{j^{'}}F^{jj^{'}}_{nn'} \psi_{n'j'}\\
&+ \delta_{n0}\sum_{j^{'}} G^{jj^{'}}\psi_{\alpha j'}+ \delta_{n\alpha} \sum_{j^{'}} G^{jj^{'}}\psi_{0j'}.
 \label{latt_eqn1}
\end{split}
\end{equation}
The terms $\sim J,\mathcal{E}_j $ on the right pertain to the main chain, 
the term $\sim F$ to the control unit, and the terms $\sim G$ represent the coupling between main chain and the control unit. The coefficients $F$ and $G$ 
are matrix elements of the electronic-vibrational coupling, given in \eref{Fs_rep}  and \eref{s_rep}, respectively.
Details on the calculation of these matrix elements for one exemplary interaction are provided in \aref{app_evibcoupling}.

In the following we employ the method of backward propagation.
Instead of specifying the vibrational state when the excitation is \emph{incoming}, as in the problem we intend to solve, we consider the problem where the \emph{outgoing} wave can be assigned a well defined vibrational 
quantum number $j_0$.
This leads to a simple outgoing boundary condition in \bref{aux_eigfn01}.
This in turn implies a more complicated superposition of vibrational states in the incoming and reflected part of the wavefunction in \bref{aux_eigfn}. 
We can finally assemble a solution that exhibits a specific incoming vibrational state as the linear combination \bref{aux_eigfn1} of these auxiliary scattering solutions.

One crucial part of solving the eigenvalue problem is obtaining the probability amplitude on the scatterer. With the regular boundary conditions, this is a difficult task in the presence of many vibrational levels. Below we illustrate the method to find the probability amplitudes on the scatterer, which is essential in solving the auxiliary equations. The particular choice of boundary conditions discussed above along with the auxiliary equations makes this easier.

We can explicitly solve the auxiliary eigenproblem \bref{aux_latt_eqn} for all the possible values of the index $j_0$. From \eref{aux_eigfn}, we know that $\psi^{(j_0)}_{0j} = \delta_{jj_0}$. 
This is the key property of the auxiliary problem that simplifies the determination of all vibrational amplitudes in the scatterer compared to the 
original problem, where in general all $\psi_{0j} $ may be nonzero.
In order to obtain the wavefunction associated with the ring, i.e., to get the quantities $\psi^{(j_0)}_{\bullet j} \equiv \{\psi^{(j_0)}_{\alpha j}, \psi^{(j_0)}_{\beta j},  \psi^{(j_0)}_{\eta j}\}$,
one can now deal with an inhomogeneous system of linear equations, which is directly obtained from \eref{aux_latt_eqn} and \eref{latt_eqn1} as 
\begin{widetext}
\begin{eqnarray}\label{ring_eqn1a}
{ }& (E - \mathcal{E}_{j} )   \psi_{\alpha j} -
  \sum_{jj^{'}} F^{jj^{'}}_{\alpha \beta} \psi_{\beta j^{'}} -  \sum_{j^{'}} F^{jj^{'}}_{\alpha \eta} \psi_{\eta j^{'}} &= \sum_{j^{'}}G^{jj^{'}}\psi_{0 j^{'}},\\  \nonumber
{ }& (E - \mathcal{E}_{j} )  \psi_{\beta j} -
  \sum_{j^{'}} F^{jj^{'}}_{\beta \alpha } \psi_{\alpha j^{'}} -  \sum_{j^{'}} F^{jj^{'}}_{\beta \eta} \psi_{\eta j^{'}}  &=0,\\\nonumber
 { }& (E - \mathcal{E}_{j} )   \psi_{\eta j} -
  \sum_{j^{'}} F^{jj^{'}}_{\eta \alpha } \psi_{\alpha j^{'}} -  \sum_{j^{'}} F^{jj^{'}}_{\eta \beta} \psi_{\beta j^{'}}  &=0.
\end{eqnarray}
\end{widetext}
Problem (\eref{ring_eqn1a}) must always possess a unique solution $\psi^{(j_0)}_{\bullet j}$  as long as the entire theoretical formulation is consistent. 
After determining the quantities $\psi^{(j_0)}_{\bullet j}$, which depend 
on all the $\psi_{0,j'}$, the amplitude $\psi^{(j_0)}_{-1, j}$ can be obtained from 
\begin{eqnarray}
J \psi_{+1, j}+ J \psi_{-1, j}+ ( \mathcal{E}_{j} - E ) \psi_{0, j} +\nonumber
 \\\sum_{j^{'}} G^{jj^{'}} \psi_{\alpha j^{'}}=0. 
\label{psi_minus1}
\end{eqnarray}
It then allows us to find the coefficient $A_j^{j_0}$ after writing down \eref{aux_eigfn} for the case $n = -1$ and inserting $\psi^{(j_0)}_{-1, 
j}$  from \eref{psi_minus1}.
 One then finds
\begin{eqnarray}
2 i A_j^{j_0} \sin{k_j} = \delta_{jj_0}e^{+ik_{j_0}}-\psi_{-1, j}.
\label{ajj0}
\end{eqnarray}
This completes the explicit solution of the eigenproblem \bref{aux_eigfn}.

\section{Electronic - vibrational Coupling} 
\label{app_evibcoupling}
 In this appendix, we derive the expressions for the components $ F_{nn'}^{jj'} $ and $ G^{jj'} $ defined in \eref{Fs_rep}  and \eref{s_rep}, respectively for the specific example of dipole-dipole interactions, where
\begin{eqnarray}
F_{nn'}(\boldsymbol{\vartheta})&=-\frac{\mu^2}{|\mathbf{r}_n(\vartheta_n) - \mathbf{r}_{n'}(\vartheta_{n'})|^3}
\label{dipdipint}
\end{eqnarray} 
and similarly
\begin{eqnarray}
G(\boldsymbol{\vartheta})&=-\frac{\mu^2}{|\mathbf{r}_0 - \mathbf{r}_{\alpha}(\vartheta_{\alpha})|^3},
\label{dipdipinta}
\end{eqnarray} 
 with transition dipole moment $\mu$ and $\mathbf{r}_n(\vartheta_n)$ the position of monomer $n$.  Other exponents for the distance dependence such as $1/r^m$ (with $m>2$) would lead to structurally similar expression for the matrix elements and hence qualitatively similar results. The same would be true for any other interactions for which the interaction between the monomers in the control unit as well as that between main chain chain and the control unit depends on the distance between the monomers.

Let $ \vartheta_{n0}$ denote the central angle of monomer $n$ on the ring 
where the vibrational potential has its minimum. The angular position of monomer $n$ is given by $\vartheta_n$, and the displacement hence defined 
as $\Delta_n$ as $ \vartheta_{n} -  \vartheta_{n0}$. The trap potential in the position representation is then:
\begin{eqnarray}
V_{trap} (\Delta_n) &= \frac{1}{2} M \omega^2 R^2 \Big(2\big(1-\cos{\Delta_n}\big)\Big)\\ &\approx  \frac{1}{2} M \omega^2 (R \Delta_n)^2.
\label{v_trap}
\end{eqnarray}
\paragraph{Calculation of $F_{nn'}^{jj'}$:\\ }
We now focus in the term $F_{nn'}$, i.e., interaction between monomers on 
the ring.
The inverse cubed distance between two monomers $n$ and $n'$ on the ring can be expressed
through their angular coordinates $\vartheta_{n}$ and $\vartheta_{n'}$ as

\begin{eqnarray}
r_{nn'}^{-3}=\frac{2^{-3/2}}{R^3}\Big(1-\cos{(\vartheta_n- \vartheta_{n'})}\Big)^{-3/2}.
\label{rnn}
\end{eqnarray}
The angular separation between monomers $n$ and $n'$ is denoted by $\vartheta_{nn'}= \vartheta_n- \vartheta_{n'}$, hence an
equidistant configuration of them corresponds to  $| \vartheta_{nn'}|= 2\pi/3$ for all pairs $n,n'$. For convenience let $\Delta_{nn'} = \Delta_n-\Delta_{n'}$ . We then can write
\begin{equation}
r_{nn'}^{-3}=\frac{2^{-3/2}}{R^3}\Big(1-\cos(\vartheta_{nn'}+\Delta_{nn'})\Big)^{-3/2}.
\label{rnn1}
\end{equation}
Assuming the displacements to be small, a Taylor-expansion of the function $f(\Delta_{nn'})=\Big(1-\cos( \vartheta_{nn'}+\Delta_{nn'})\Big)^{-3/2}$ up to first order around $\Delta_{nn'} = 0$ gives
\begin{align}
f(0) &= (1 - \cos{\vartheta_{nn'}})^{-3/2} & \equiv F_0^{nn'}, \label{fs1} \\ 
f'(0) &= -\frac{3}{2} (1 - \cos{\vartheta_{nn'}})^{-5/2}\sin{\vartheta_{nn'}}& \equiv F_1^{nn'}.
\label{fs2}
\end{align}
Hence the approximate inverse cubed distance is
\begin{eqnarray}
r_{nn'}^{-3}= \frac{2^{-3/2}}{R^3}(F_0^{nn'}+F_1^{nn'}\Delta_{nn'}).
\label{rnn2}
\end{eqnarray}
This leads to
\begin{eqnarray}
F_{nn'}^{jj^{'}} &=& ~_{vib}\bra{\Phi_j}F_{nn'}(\boldsymbol{\vartheta})\ket{\Phi_{j^{'}}}_{vib} \nonumber \\
&=& \delta_{j_zj^{'}_z}\int\int d(R\Delta_n) d(R\Delta_{n'}) \Phi_{j_x}(R\Delta_n) \Phi_{j_y}(R\Delta_{n'}) \nonumber 
\\  &&(F_0^{nn'}+F_1^{nn'} \Delta_{nn'})\Phi_{j^{'}_x}(R\Delta_n) \Phi_{j^{'}_y}(R\Delta_{n'}).
\label{fnn}
\end{eqnarray}
Here $z$ denotes the monomer index on the ring that should be neither $n$ 
nor $n'$, which is uniquely determined since for this term we also require $n\ne n'$ and there are only three monomers in total. $j \equiv \{j_\alpha,j_\beta,j_\eta\}$ represents the vibrational state of each monomer in 
ring and $x, y, z \in \{\alpha,\beta,\eta\} $\textbackslash$ \{n, n'\}$. For example, if $n=\alpha$ and $n'=\eta$, then $j_\alpha \rightarrow j_x$ , $ j_\beta \rightarrow j_z$ and $j_\eta \rightarrow j_y$ and hence $j$ and $j^{'}$ can be written as $j \equiv \{j_x,j_z,j_y\}$ and $j^{'} \equiv \{j^{'}_x,j^{'}_z,j^{'}_y\}$. Further,  the complex conjugation of the eigenfunctions is omitted since those are real. Finally, since  a finite number of modes are included, the integration can be formally extended from $-\infty$ to $+\infty$.  Hence,
\begin{eqnarray}
F_{nn'}^{jj^{'}} =  \delta_{j_zj^{'}_z}(I_0+I_1),
\label{fnn1}
\end{eqnarray}
where
\begin{eqnarray}
I_0 &=& F_0^{nn'} \delta_{j_xj^{'}_x} \delta_{j_yj^{'}_y}, \\
 I_1 &=& F_1^{nn'}\int\int d(R\Delta_n) d(R\Delta_{n'}) \Phi_{j_x}(R\Delta_n) \Phi_{j_y}(R\Delta_{n'}) \nonumber \\&&( \Delta_{n} - \Delta_{n'})\Phi_{j^{'}_x}(R\Delta_n) \Phi_{j^{'}_y}(R\Delta_{n'}) \\
 {  } &=& F_1^{nn'}\int d(R\Delta_n) \Phi_{j_x}(R\Delta_n) \Delta_{n} \Phi_{j^{'}_x}(R\Delta_n) - \nonumber 
 \\&& \int  d(R\Delta_{n'}) \Phi_{j_y}(R\Delta_{n'})  \Delta_{n'} \Phi_{j^{'}_y}(R\Delta_{n'}) .
\label{fnn2}
\end{eqnarray}
Using the explicit form of the eigenfunction in terms of Hermite polynomials $H_j(x)$, the first integral in \eref{fnn2} becomes
\begin{widetext}
\begin{align}
&\int d(R\Delta_n) \Phi_{j_x}(R\Delta_n) \Delta_{n} \Phi_{j^{'}_x}(R\Delta_n) = \\
&\sqrt{\frac{M\omega}{\pi \hbar} }\frac{1}{\sqrt{2^{j_x+j^{'}_x}j_x!j^{'}_x!}} \int d(R\Delta_n)  H_{j_x}\left(\sqrt{\frac{M\omega}{\hbar}}R\Delta_n\right) \Delta_n  H_{j^{'}_x}\left(\sqrt{\frac{M\omega}{\hbar}}R\Delta_n\right) \exp\left( -\frac{M\omega}{\hbar}R^2\Delta_n^2\right)\\
&=  \frac{1}{2R}\sqrt{\frac{\hbar}{\pi M \omega}}\frac{1}{\sqrt{2^{j_x+j^{'}_x}j_x!j^{'}_x!}} \int dX H_{j_x}(X) H_{j^{'}_x}(X) H_1(X)\exp\left(-X^2\right),
\label{fnn3}
\end{align}
\end{widetext}
where $X= \sqrt{\frac{M\omega}{\hbar} } R\Delta_n$ and also $X = H_1 (X)/2$.

This integral vanishes whenever $j_x + j^{'}_x +1$ is odd, since in this case the integrand is odd and the integration interval symmetric. Moreover, one needs $|j_x - j^{'}_x| = 1$ since after Taylor expansion to first order the harmonic oscillator ladder operators couple only adjacent vibrational states. For even $j_x + j^{'}_x +1$ we can obtain \cite{gradshteyn2014table}
\begin{eqnarray}
 \int dX H_{j_x}(X) &H_{j^{'}_x}(X) H_1(X)\exp\left(-X^2\right) \nonumber
  \\ &= \frac{2^{s_x}\sqrt{\pi}j_x ! j^{'}_x!}{(s_x -j_x)!(s_x-j^{'}_x)!(s_x-1)!},
\label{fnn4}
\end{eqnarray}
with $s_x = \frac{j_x + j^{'}_x +1}{2}$.
Since $|j_x - j^{'}_x| = 1$, $(s_x -j_x)!(s_x-j^{'}_x)!(s_x-1)! = \left( \frac{j_x + j^{'}_x  - 1}{2}\right)!$. Now let 
\begin{equation}
\Gamma(v,w) = \frac{\sqrt{v!w!}}{\left(\frac{v+w-1}{2}\right)!}.
\end{equation}
 Then $I_1$ can be written as, 
\begin{eqnarray}
 I_1 = F_1^{nn'} \frac{1}{R} \sqrt{\frac{\hbar}{2 M \omega}} \left(\Gamma(j_x,j^{'}_x)-\Gamma(j_y,j^{'}_y)\right).
\label{fnn5}
\end{eqnarray}
Thus 
\begin{align}
F_{nn'}^{jj^{'}} =& -\frac{\mu^2}{(\sqrt{2}R)^3} \delta_{j_zj^{'}_z}\Bigg( \Bigg. F_0^{nn'} \delta_{j_xj^{'}_x} \delta_{j_yj^{'}_y} 
 \nonumber  \\
+& \frac{F_1^{nn'}}{R} \sqrt{\frac{\hbar}{2 M \omega}} \left(\Gamma(j_x,j^{'}_x)-\Gamma(j_y,j^{'}_y)\right)\Bigg. \Bigg),
\label{fnn6}
\end{align}
with $F_k^{nn'}$ defined in \eref{fs1} and \eref{fs2}. Finally, since $|j_x - j^{'}_x| = 1$ and $|j_y - j^{'}_y| = 1$, we can write
\begin{align}
&\Gamma(j_x,j^{'}_x)-\Gamma(j_y,j^{'}_y) = \sqrt{j_x+1}\delta_{j^{'}_x,j_x+1} + \sqrt{j^{'}_x+1}\delta_{j^{'}_x+1,j_x} \nonumber \\&- \sqrt{j_y+1}\delta_{j^{'}_y,j_y+1}- \sqrt{j^{'}_y+1}\delta_{j^{'}_y+1,j_y}.
\label{gees}
\end{align}

\paragraph{Calculation of $G^{jj'}$:\\ }
Now consider the matrix elements of $\hat{H}_{SC}$. The position of the trap $\alpha$ is at $0^o$, and so the inverse cubed distance can, for small $\Delta_{\alpha}$, be approximated by,
\begin{eqnarray}
r_{\alpha,0}^{-3} \approx ((d-R)^3+R^2\sin^2{\Delta_\alpha})^{-3/2}.
\label{ralpha}
\end{eqnarray}
However, in the Taylor-expansion of the right hand side around $\Delta_\alpha = 0$, the first non-vanishing term beyond the zeroth-order is $\sim \Delta_\alpha^2$, so that one may as well directly approximate $r^{-3} \approx (d - R)^{-3}$. This leads to,

\begin{eqnarray}
G^{jj^{'}} = -\frac{\mu^2}{(d-R)^3}\delta_{jj^{'}}\delta_{0,\alpha}.
\label{Gfinal}
\end{eqnarray}

\section{Time dependent Schr\"odinger equation (TDSE)} 
\label{app_TDSE}

The quantum dynamics of the system is governed by the time dependent Schr\"{o}dinger equation

\begin{eqnarray}
i \hbar \frac{d}{dt} \ket{\psi(t)} = \hat{H} \ket{\psi(t)},
\label{lat_eqn}
\end{eqnarray}
where $ \hat{H}$ is the full Hamiltonian of the system. Explicitly writing the equation for the state \bref{entangled_state}, we get
\begin{equation}
\begin{split}
i\hbar \dot{\psi}_{n j} &= (1 - (\delta_{n\alpha}+\delta_{n\beta}+\delta_{n\eta}))(J \psi_{n+1, j}+\\&+J\psi_{n-1, j})+ \mathcal{E}_j \psi_{nj}\\
&+ (\delta_{n\alpha}+\delta_{n\beta}+\delta_{n\eta})   \sum_{n'\in(\alpha,\beta,\eta)}^{n'\ne n} \sum_{j^{'}}F^{jj^{'}}_{nn'} \psi_{n' j^{'}}\\
&+ \delta_{n0}\sum_{j^{'}} G^{jj^{'}}\psi_{\alpha j^{'}} + \delta_{n\alpha} \sum_{j^{'}} G^{jj^{'}}\psi_{0 j^{'}}.
 \label{latt_eqn_td}
\end{split}
\end{equation}
A Gaussian wave packet far left of the side-unit, representing the incoming excitation, is our initial condition for solving \eref{lat_eqn}. Initially, the monomers on the circle are assumed to be 
in the vibrational ground state. The excitation propagates freely towards 
the right as long as the $n=0$ site remains unpopulated. During this pre-collision time interval, vibrational degrees of freedom of the ring monomers are unaffected by the incoming excitation and hence remain in the original stationary state. The dynamical time evolution remains governed solely by the first term ($\hat{H}_S$) in the Hamiltonian. The situation changes significantly as soon as the excitation reaches $n=0$ site. The other terms in the Hamiltonian 
become important and a complicated vibrational quantum dynamics takes place until the excitation completely leaves the scattering region. The electronic-vibrational coupling term in the Hamiltonian could take the monomer to the higher vibrational states, and thereby influence the excitation transport in the main chain. The post-collision dynamics is again essentially governed by $\hat{H}_S$. 

The excitation probability of the monomers on the left or the right of the Fano defect obtained during the pre-collision and post-collision period can be used to define a 
transmission and 
reflection coefficient. In addition, the contribution from each channel to the transmission coefficient can also be calculated from the dynamics by projecting the spatial wavefunction onto harmonic oscillator states after the scattering.

A Gaussian wavepacket has an energy width arising from spatial localisation within $\Delta$ given by
\begin{equation}
\Delta E_{packet} \approx \frac{2\hbar^2 k}{m \Delta},
\label{gauss}
\end{equation}
where the mass $m$ in a tight binding contest can be expressed as $m = \frac{1}{\frac{1}{\hbar^2}\frac{d^2E}{dk^2}}$ and $k$ is the wavevector.

\begin{figure}[htb]
\includegraphics[width=0.8\columnwidth]{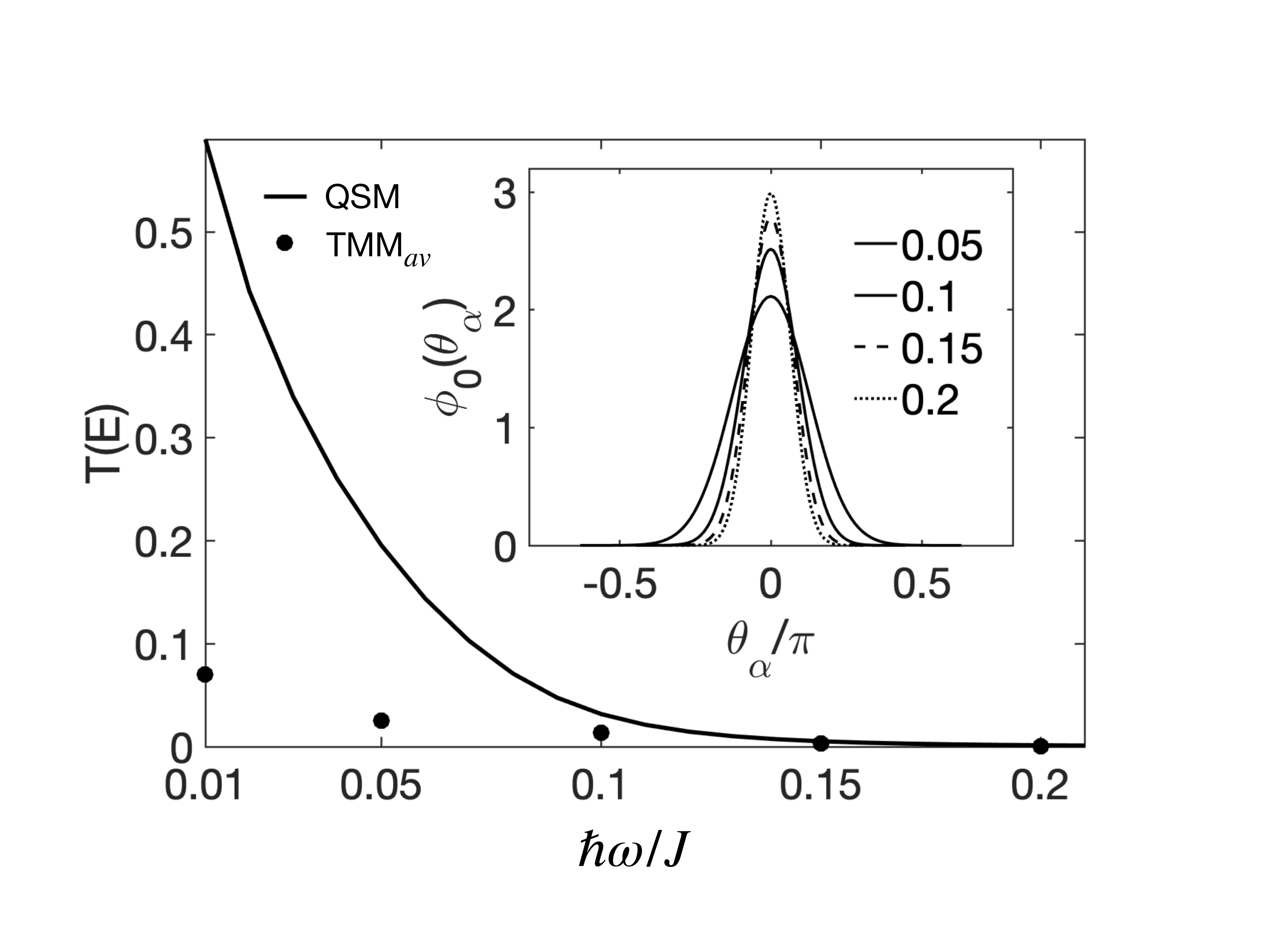}
\caption{ Transmission as a function of frequency of vibration at the incoming excitation energy $E=J$. The solid line shows the transmission obtained from the QSM. 
The black dots show the transmission obtained using the static model given by \eref{T_from_static_model}.  Inset: The spacial width of the initial 
ground state considered for the vibrating monomer $\alpha$ for different vibrational frequencies with energy $\hbar\omega/J$.}
\label{TvsOmega}
\end{figure}
%
\section{Transfer Matrix Method} 
\label{tmm}

In this appendix we consider the arrangement of \fref{geometry} for static sites. With the full state  $\ket{\Psi}=\sum_n \psi_n \ket{n}$ the time independent Schr\"{o}dinger equation for
quantum scattering
of incoming waves on the CU turns into
%
\begin{align} \nonumber
 E \psi_n &=  J \psi_{n+1}+J \psi_{n-1}  + G \psi_\alpha \delta_{n,0},\\ \nonumber
E \psi_\alpha &=
 F_{\alpha \beta} \psi_\beta +   F_{\alpha \eta} \psi_\eta +
 G \psi_0,    \\ \nonumber
E \psi_\beta &=
 F _{\beta \alpha } \psi_\alpha +  F_{\beta \eta} \psi_\eta,  \\
E \psi_\eta &=
 F_{\eta \alpha } \psi_\alpha +   F_{\eta \beta} \psi_\beta\,,
\label{latt_eqn1a}
\end{align}
where we have used the short-hand notation $G=G^{00}$, with $G^{ij'}$ defined in \bref{s_rep}.
Rearranging \eqref{latt_eqn1a} gives
\begin{align} 
 E \psi_n =  J \psi_{n+1} + J \psi_{n-1}  + \frac{D_2}{D_3}G^2 \psi_0 \delta_{n,0}
\label{latt_eqn1b}
\end{align}
with $D_2 = \mbox{det}(E \mathbb{1} - H_2), ~D_3 = \mbox{det}(E \mathbb{1}  - H_3)$, where
\begin{align} 
H_3 = \begin{bmatrix} 
0 & F_{\alpha\beta} & F_{\alpha\eta} \\
 F_{\beta\alpha}  & 0 &  F_{\beta\eta}  \\
  F_{\eta\alpha}  & F_{\eta\beta}  & 0
\end{bmatrix},\:\:
H_2 = \begin{bmatrix} 
0 & F_{\beta\eta} \\
 F_{\eta\beta}  & 0 
 \end{bmatrix}
\label{latt_eqn1c}
\end{align}
are the dipole-dipole Hamiltonian of the CU ($H_3$) and of the reduced CU without the entrance site  ($H_2$). We see in \bref{latt_eqn1b}
that the CU acts like  a localized defect on site $n=0$. The strength of the effective defect potential 
\begin{equation} 
\sub{V}{eff} = D_2G^2/D_3
\label{veff}
\end{equation}
depends on the energy of the incoming excitation through the $D_{j}$. If it matches one of the eigenenergies of the side-unit, $D_3=0$ and the diverging effective scattering potential leads to a total reflection of the incoming wave.
In contrast, when for energies resonant on an eigenenergies of the side-unit minus entrance site, $D_2=0$,  the effective scattering potential vanishes and we have perfect transmission \cite{kamenetskii2018fano}.

To explicitly evaluate the reflection and transmission coefficient at other energies, we make the usual Ansatz
\begin{eqnarray} 
 &\psi_n &= i_\mathrm{in} e^{+ik n}+r_\mathrm{o} e^{-ik n}  \hskip0.8cm 
n<0, \label{psi_b_cond2a}\\
 &\psi_n &= t_\mathrm{o} e^{+ik n} \hskip2.4cm n>0.
\label{psi_b_cond2b}
\end{eqnarray}
The equation \bref{latt_eqn1b} can be written in the form
\begin{equation} 
\begin{bmatrix} 
\psi_{n+1} \\
 \psi_n 
  \end{bmatrix} = 
T_n \begin{bmatrix} 
\psi_n \\
 \psi_{n-1} 
 \end{bmatrix},
\label{tmm_1}
\end{equation}
with a transfer matrix
\begin{equation} 
T_n = \begin{bmatrix} 
\frac{E}{J} - \frac{\sub{V}{eff}}{J} \delta_{n,0} & -1\\
 1 & 0
\end{bmatrix}.
\label{t_n}
\end{equation}
Thus,
 \begin{equation} 
\begin{bmatrix} 
\psi_{n+1} \\
 \psi_n 
  \end{bmatrix} = 
P_n \begin{bmatrix} 
\psi_{-n} \\
 \psi_{-n-1} 
 \end{bmatrix}
\label{tmm_2} , 
\end{equation}
with  $P_n = T_nT_{n-1}\dots T_{-n}$.  Using the boundary conditions in 
\eref{psi_b_cond2a} and \bref{psi_b_cond2b},  the transmission coefficient $T=|t_\mathrm{o}|^2/|i_\mathrm{in}|^2$ can be obtained from (\eref{tmm_2}) as \cite{kamenetskii2018fano},
\begin{equation}
T (E)= \frac{4J^2 - E^2}{4J^2 - E^2+\sub{V}{eff}^2} .
\label{trans_main}
\end{equation}
with $\sub{V}{eff}$ given in \eref{veff}.

\section{Static transmission averages} 
\label{static}

When approaching smaller vibrational frequencies $\omega$, the zero-point 
width $\sigma=\sqrt{\hbar/m/\omega}$ of the vibrating monomer increases. 
Since we can understand transmission for immobile monomers using the TMM discussed in \sref{tmm}, we can attempt to make contact with those calculation by taking the 
transmission from the TMM for a fixed angle $\theta$ of the vibrating monomer $\alpha$, let that be $T_{TMM}(\theta)$, and then averaging it according to
\begin{eqnarray}
T=\int T_{TMM}(\theta)p(\theta)d\theta
\label{T_from_static_model}
\end{eqnarray}
over the position distribution $p(\theta)$ in the harmonic oscillator ground-state, shown in the inset of \fref{TvsOmega}. However when applied to 
the scenario of e.g.~\fref{trans_vib} this provides transmission of at most $T\approx 0.08$ near $E=J$, clearly not capturing the essential physics which shows a much more prominent increase of transmission, see  \fref{TvsOmega}. The underlying resonance peak shifts away from $E=J$ as the frequency increases and thus the excitation transport increases at lower vibrational frequencies.
\end{document}